\documentclass[prd,twocolumn,showpacs,preprintnumbers,amsmath,amssymb]{revtex4}
\usepackage{graphicx}
\usepackage{bm}

\newcommand{\beq}{\begin{equation}}
\newcommand{\eeq}{\end{equation}}
\newcommand{\beqs}{\begin{eqnarray}}
\newcommand{\eeqs}{\end{eqnarray}}

\newcommand{\gsim}{\mathrel{\raisebox{-
.6ex}{$\stackrel{\textstyle>}{\sim}$}}}

\begin{document}

\title{Technifermion Representations and Precision Electroweak Constraints} 

\author{Neil D. Christensen}
\author{Robert Shrock}

\affiliation{C.N. Yang Institute for Theoretical Physics \\
State University of New York \\
Stony Brook, NY 11794}

\begin{abstract}

We discuss the selection of fermion representations in technicolor models with
a view toward minimizing technicolor contributions to the precision electroweak
$S$ parameter. We present and analyze models that involve one technifermion
SU(2)$_L$ doublet with standard-model singlet technifermion sectors that lead
to walking behavior, which further reduces $S$.  We also consider models that
have technifermions in higher-dimensional representations and study embeddings
in extended technicolor theories.

\end{abstract}

\maketitle

\section{Introduction}

It is possible that electroweak symmetry breaking occurs as the result of the
existence of a new, asymptotically free, vectorial gauge interaction,
generically called technicolor (TC) \cite{tc}, which becomes strongly coupled
at a scale $\Lambda_{TC}$ of several hundred GeV, producing a bilinear
technifermion condensate with weak isospin $I=1/2$ and weak hypercharge $Y=1$.
To communicate the electroweak symmetry breaking to the standard-model
(technisinglet) fermions and to give them masses, one embeds technicolor in a
larger, extended technicolor (ETC) theory \cite{etc} (reviews include
\cite{etcrev}).  Technicolor theories produce corrections to precisely measured
electroweak quantities, in particular, to the $W$ and $Z$ propagators (called
oblique corrections), and are stringently constrained by the requirement that
these modifications not exceed experimental limits.  Here we analyze
technifermion representations from the viewpoint of minimizing these
technicolor corrections, in particular, the $S$ parameter.  We construct models
that can accomplish this, using (a) an SU($N_{TC}$) gauge group with the
minimal nonabelian value, $N_{TC}=2$; (b) a minimal standard-model
(SM)-nonsinglet sector consisting of technifermions that transform as a doublet
under weak isospin SU(2)$_L$ and also a doublet under SU(2)$_{TC}$; and (c) a
SM-singlet technifermion sector that produces walking behavior, which further
reduces $S$. We also consider models with technifermions in higher-dimensional
representations of the TC group, and study embeddings of both types of
technicolor models in extended technicolor theories.

\section{Some Basics}

In a minimal technicolor model with gauge group $G_{TC}$ the technifermions
transform under $G_{TC} \times {\rm SU}(3)_c \times {\rm SU}(2)_L \times {\rm
U}(1)_Y$ as
\beq
{F^{1/2} \choose F^{-1/2}}_L \ : \ ({\cal R}_{TC},1,2)_{0,L}, \quad 
F^{\pm 1/2}_R \ : \ \ ({\cal R}_{TC},1,1)_{\pm 1,R}, 
\label{1d}
\eeq
where ${\cal R}_{TC}$ denotes the representation of $G_{TC}$, and the
superscripts (subscripts) denote electric charge (weak hypercharge $Y$
and chirality), respectively.  The $Y$ values in eq. (\ref{1d}) are determined
by the requirement of no SU(2)$^2_L$U(1)$_Y$ or U(1)$^3_Y$ gauge anomalies.
Here we take $G_{TC}={\rm SU}(N_{TC})$.  Most studies have chosen for ${\cal
R}_{TC}$ the simplest nontrivial possibility, namely, the fundamental
representation, but there have also been studies with higher-dimensional
technifermion representations \cite{higherrep}-\cite{sann3}.  At certain points
below it will be useful to compare predictions of TC theories based on
eq. (\ref{1d}) with those in which the technifermions transform as a SM
family; some recent works on this latter type of model are
Refs. \cite{at94}-\cite{lane}.

Technicolor corrections to the $W$ and $Z$ propagators are summarized in terms
of the $S$, $T$, and $U$ parameters \cite{pt,scalc1,scalc2} (for reviews, see
\cite{pdg,lepewwg}).  Of these, the $S$ and $T$ parameters provide the most
important constraints on technicolor.  We denote the technicolor contributions
to $S$ and $T$ as $(\Delta S)^{(TC)}$ and $(\Delta T)^{(TC)}$.  The $T$
parameter measures corrections, from new physics (NP) beyond the standard
model, to the custodial symmetry relation $\rho = 1$, where $\rho =
m_W^2/(m_Z^2 \cos^2 \theta_W)$ and $\Delta \rho^{(NP)} = \alpha_{em}(m_Z) T$.
Since the SM gauge interactions are small at the scale $\Lambda_{TC}$,
technifermion condensates can naturally produce nearly degenerate dynamical
masses for SM-nonsinglet technifermions with weak $I_3 = \pm 1/2$, preserving
approximate custodial symmetry and yielding an acceptably small $|(\Delta
T)^{(TC)}|$. One of the tasks that ETC theories take on is then how to explain
the large $t$-$b$ mass splitting while maintaining a small $|(\Delta
T)^{(TC)}|$. One-family ETC models using relatively conjugate ETC
representations for left- and right-handed $Q=-1/3$ quarks can account for this
$m_t - m_b$ splitting without excessive contributions to $T$, but have problems
with flavor-changing neutral currents (FCNC) \cite{ckm,kt}.  (In contrast, for
models with vectorial ETC representations, it was shown in Refs. \cite{ckm,kt}
that FCNC constraints, in particular from $K^0 - \bar K^0$ mixing, are not as
serious as had been thought previously.)

The $S$ parameter measures heavy-particle contributions to the $Z$ self-energy
via the term $4 s_W^2 c_W^2 \alpha^{-1}_{em}(m_Z)[\Pi^{(NP)}_{ZZ}(m_Z^2) -
\Pi^{(NP)}_{ZZ}(0)]/m_Z^2$, where $s_W^2 = 1-c_W^2 = \sin^2\theta_W$, evaluated
at $m_Z$ (see \cite{pdg} for details). Here we shall focus on the minimization
of $(\Delta S)^{(TC)}$, commenting on $(\Delta T)^{(TC)}$ briefly below.
Global fits to data yield allowed regions in $(S,T)$ depending on a reference
value of the SM Higgs mass, $m_{H,ref.}$.  The comparison of these with a
technicolor theory is complicated by the fact that technicolor has no
fundamental Higgs field; sometimes one formally uses $m_{H,ref.} \sim 1$ TeV
for a rough estimate, since the SM with $m_H \sim 1$ TeV has strong
longitudinal vector boson scattering, as does technicolor. However, this may
involve some double-counting when one also includes contributions to $S$ from
technifermions, whose interactions and bound states (e.g., techni-vector
mesons) are responsible for the strong $W^+_L W^-_L$ and $Z_LZ_L$ scattering in
a technicolor framework.  The current allowed region in $(S,T)$ \cite{lepewwg}
disfavors values of $S \gsim 0.2$ and $|T| \gsim 0.2$.

For fermions comprising an SU(2)$_L$ doublet, plus right-handed SU(2)$_L$
singlets, which have degenerate masses $m_F$ satisfying $(2m_F/m_Z)^2 \gg 1$
and are weakly interacting, the well-known one-loop contribution to $S$ is
$N_D/(6\pi)$ (independent of $Y$).  Since technifermions are strongly
interacting on the scale $m_Z$ used in the definition of $S$, it is of
questionable validity to try to apply perturbation theory to calculate $(\Delta
S)^{(TC)}$. Nevertheless, the estimate of $(\Delta S)^{(TC)}$ based on the
perturbative one-loop contribution of the technifermions is often used as an
approximate guide.  Because the technifermions have dynamical masses
$\Sigma_{TC}$ that satisfy $(2\Sigma_{TC}/m_Z)^2 \gg 1$ and which, moreover,
are naturally approximately degenerate, it follows that a perturbative estimate
is $(\Delta S)^{(TC)}_{pert.}  = N_D/(6\pi)$, where $N_D$ denotes the total
number of new technifermion SU(2)$_L$ doublets.  For the model of
eq. (\ref{1d}), commonly called the ``one-doublet'' TC model, this total number
is $N_D=dim({\cal R}_{TC})$, while for a one-family TC model, $N_D=(N_c+1) \,
dim({\cal R}_{TC}) = 4 \, dim({\cal R}_{TC})$ (where $N_c=3$ colors).
Therefore, to minimize $(\Delta S)^{(TC)}$, one can reduce $N_{TC}$ to its
minimal nonabelian value, $N_{TC}=2$ and ${\cal R}_{TC}$ to its smallest
nontrivial possibility, viz., the fundamental ($fund.$) representation.  With
these choices, the TC model of eq. (\ref{1d}) yields
\beq (\Delta S)^{(TC)}_{pert.}=\frac{1}{3\pi} \ , \quad {\rm for} \ \ N_{TC}=2,
\ \ {\cal R}_{TC}=fund.,
\label{deltas_1d}
\eeq
while $(\Delta S)^{(TC)}_{pert.} = 4/(3\pi)$ for the one-family TC model.
Higher ${\cal R}_{TC}$ are discussed below.  Another advantage of the model of
eq. (\ref{1d}) is that (for general $N_{TC}$) all of the three Nambu-Goldstone
bosons (NGB's) that arise due to the formation of technicondensates are
absorbed to make the $W^\pm$ and $Z$ massive so that there are no problems with
unwanted (pseudo) NGB's.

An important property of modern technicolor theories is a TC gauge coupling
that runs slowly (``walks'') over a certain energy interval extending from
$\Lambda_{TC}$ to a higher ETC scale, $\Lambda_w$ \cite{wtc,chiralpt}.  Walking
technicolor (WTC) occurs naturally if the TC gauge coupling has an approximate
infrared-stable fixed point (zero of the beta function) $\alpha_{TC,IR}$ which
is slightly larger than the critical value $\alpha_{TC,c}$ for technifermion
condensate formation.  In such a theory, as the energy scale $\mu$ decreases
from large values, $\alpha_{TC}$ increases, but its rate of increase, given by
$-\beta$, decreases as $\alpha_{TC}$ approaches the zero at $\alpha_{TC,IR}$.
Hence, over an extended energy interval, $\alpha_{TC}$ is O(1) but slowly
varying.  This is accompanied by a large anomalous dimension
$\gamma \simeq 1$ for the bilinear technifermion operator $\bar F F$, resulting
in the enhancement of SM fermion masses by the factor $\eta =
\exp[\int_{\Lambda_{TC}}^{\Lambda_w} (d\mu/\mu) \gamma(\alpha(\mu))] \simeq
\Lambda_w/\Lambda_{TC}$ and also enhancement of pseudo-Nambu-Goldstone boson
masses.  In a non-walking scaled-up QCD type of technicolor theory,
spectral-function methods yield $(\Delta S)^{(TC)} \simeq 0.1 N_D \simeq
2(\Delta S)^{(TC)}_{pert.}$ \cite{pt}.  Nonperturbative estimates of $(\Delta
S)^{(TC)}$ in WTC models show that it is reduced relative to nonwalking TC
\cite{scalc2}, clearly a desirable feature.

An analysis of the beta function of the one-family technicolor model with
$N_{TC}=2$ and (vectorially coupled) technifermions transforming according to
the fundamental representation, i.e., techni-isospin $I_{TC}=1/2$, suggests
that, with its $N_w(N_c+1)=8$ technifermions, it can plausibly exhibit walking
behavior \cite{wtc,chiralpt} (cf. eq. (\ref{nfca})). The value $N_{TC}=2$ has
been used for many studies of one-family ETC models \cite{at94}-\cite{kt} and
also has the advantage that it makes possible a mechanism to obtain light
neutrino masses \cite{nt}.  In contrast, although the technicolor model with
the minimal SM-nonsinglet technifermion sector of eq. (\ref{1d}) with ${\cal
R}_{TC}=fund.$ yields a relatively small value of $(\Delta S)^{(TC)}_{pert.}$,
especially for $N_{TC}=2$, its two-loop beta function does not have a
perturbative IR fixed point or resultant walking behavior.  Hence, it may have
difficulty producing sufficiently large SM fermion masses, in particular,
$m_t$.

\section{Minimal Technicolor Models with Walking} 

There is thus motivation for constructing technicolor models that have small
values of $(\Delta S)^{(TC)}_{pert.}$ and also have walking behavior to reduce
the full (nonperturbatively calculated) $(\Delta S)^{(TC)}$.  We proceed to do
this.  The idea is to use the model of eq. (\ref{1d}) with the smallest
nonabelian value, $N_{TC}=2$, and the minimal choice, ${\cal R}_{TC}=fund.$,
i.e., $I_{TC}=1/2$, together with a SM-singlet, TC-nonsinglet fermion sector
that produces the walking.  As noted above, this theory has walking behavior
for eight $I_{TC}=1/2$ technifermions. Since there are already $N_w=2$ such
technifermions from the SM-nonsinglet sector given in eq. (\ref{1d}), we use
six SM-singlet, $I_{TC}=1/2$ technifermions.  These should transform
nontrivially under a second vectorial gauge symmetry, denoted metacolor, which
becomes strongly coupled on a scale $\Lambda_{MC} \simeq \Lambda_{TC}$.  The
reason for having the SM-singlet technifermions be nonsinglets under metacolor
rather than just consisting of the set $\psi^\tau_{p,R}$ with $p=1,..,12$
(where, without loss of generality, we write SM-singlet fermion fields as
right-handed and use the fact that 12 such fermions are equivalent to six
Dirac fermions for SU(2))), is that, in the approximation that one neglects SM
gauge interactions, which are small at the scale $\Lambda_{TC}$, relative to TC
gauge interactions, the latter model would have a global chiral symmetry which
would be spontaneously broken by the formation of the technicondensates.  The
subset of the resultant pseudo-Nambu-Goldstone bosons (PNGB's) corresponding to
global transformations between the $\psi^\tau_{p,R}$ and $F^{\pm 1/2 \ \tau}_R$
would be color-singlets with electric charges $\pm 1/2$ and would gain masses
of order $e \, \Lambda_{TC} \sim 100$ GeV due to the explicit breaking of the
global chiral invariance by electroweak interactions.  These masses are close
enough to current experimental limits on new charged leptons, e.g. from LEP, to
disfavor such a model.

There are several possibilities for the SM-singlet technifermion
representations under metacolor.  We shall discuss two in particular.  Let us
assume that the metacolor gauge group is SU(2)$_{MC}$.  Then under ${\rm
SU}(2)_{TC} \times {\rm SU}(2)_{MC}$ these representations could be
\begin{enumerate}

\item 

six copies of (2,2), denoted $\zeta^{\tau \alpha}_{p,R}$, $p=1,..,6$,  or 

\item 

four copies of (2,3), denoted $\vec{\zeta}^\tau_{p,R}$, $p=1,..,4$, 

\end{enumerate}
where $\tau$ and $\alpha$ are TC and MC indices, $\vec{\zeta}$ refers to the MC
isovector, and $p$ is the copy number.  With the strongly coupled
metacolor, even neglecting SM gauge interactions, a global transformation of
the form $\zeta^{\tau \alpha}_{p,R} \leftrightarrow F^{\pm \ \tau}_R$ or
$\vec{\zeta}^\tau_{p,R} \leftrightarrow F^{\pm \ \tau}_R$ is not a symmetry
of the model, and hence there are no problematic light electrically charged
PNGB's.  The masses generated for the charge $q = \pm 1/2$ PNGB's are of order
$\Lambda_{MC} \simeq 300$ GeV, since the MC gauge coupling is O(1); these
masses should be sufficiently high to agree with experimental limits.

We thus envision the following properties for these models.  As the energy
scale $\mu$ decreases from large values, $\alpha_{TC}$ increases but remains at
a large O(1) value throughout a substantial interval because of the walking. 
As $\mu$ approaches the comparable scales $\Lambda_{TC} \simeq \Lambda_{MC}$,
the combined attractive TC and MC interactions lead to formation of the
condensates
\beq
\langle \epsilon_{\tau \tau'} \epsilon_{\alpha \alpha'} \zeta^{\tau \alpha \
T}_{p,R} C \zeta^{\tau' \alpha'}_{p',R}\rangle 
\label{m1condense}
\eeq
in model (i) and 
\beq
\langle \epsilon_{\tau \tau'} \vec{\zeta}^{\tau \ T}_{p,R} C \cdot 
\vec{\zeta}^{\tau'}_{p',R}\rangle
\label{m2condense}
\eeq
in model (ii).  At the slightly lower scale $\Lambda_{TC}$ the technifermion
condensates $\langle \bar F F \rangle$ form.  These models yield the
appealingly small perturbative estimate (\ref{deltas_1d}) together with
walking behavior that reduces the full (nonperturbatively calculated) $(\Delta
S)^{(TC)}$ relative to its value in a nonwalking theory.

\section{Embedding of Minimal Technicolor Model in ETC}

We next discuss embedding our SU(2)$_{TC}$ models, presented in the previous
section, in an ETC theory.  We shall give some formulas for arbitrary $N_{TC}$
to show their general structure.  One possible embedding is to use the gauge
group SU($N_{ETC}$) with
\beq
N_{ETC}=N_{TC}+N_{gen.}(N_c+1)=N_{TC}+12
\label{netc1d}
\eeq
(where the number of SM fermion generations $N_{gen.}=3$) and to assign the
left-handed SM-nonsinglet technifermions with weak $I_3=\pm 1/2$ to multiplets
containing the SM fermions with the same value of $I_3$:
\beqs
& &  (F^{1/2 \ \tau},u^{aj}, \nu^j)_\chi, \cr\cr
& & (F^{-1/2 \ \tau},d^{aj}, e^j)_\chi, \quad \chi=L,R, 
\label{1detc}
\eeqs
where $\tau$, $a$, $j$, and $\chi$ denote technicolor, color, generational
indices, and chirality, respectively, and we use the compact notation
$(u^{a1},u^{a2},u^{a3}) \equiv (u^a,c^a,t^a)$, $(d^{a1},d^{a2},d^{a3}) \equiv
(d^a,s^a,b^a)$, $(e^1,e^2,e^3) \equiv (e,\mu,\tau)$, etc.  Here, 
\beq
{\rm SU}(N_{ETC}) \supset {\rm SU}(N_{TC}) \times {\rm SU}(3)_{gen.} \times 
{\rm SU}(4)_{PS}
\label{etcsubgroups}
\eeq
where the Pati-Salam SU(4)$_{PS}$ group \cite{ps} contains, as a maximal
subgroup, ${\rm SU}(3)_c \times {\rm U}(1)_{B-L}$, with $B$ and $L$ denoting
baryon and lepton number.  Hence, $[G_{ETC},G_{SM}] \ne 0$.  The left-handed
fields form the SU(2)$_L$ doublets ${F^{1/2 \ \tau} \choose F^{-1/2 \
\tau}}_L$, ${u^{aj} \choose d^{bj}}_L$, and ${\nu^j \choose e^j}_L$.  Owing to
the ``horizontal'' structure of the ETC multiplets in eq. (\ref{1detc}), the
ETC group does not include SU(2)$_L$ or (if one chooses to gauge this)
SU(2)$_R$, and
\beq
[{\rm SU}(N_{ETC}), \, {\rm SU}(2)_{L,R}]=0 \ .  
\label{etcsu2lr}
\eeq
It follows that SU($N_{ETC}$) does not contain U(1)$_{em}$ or U(1)$_Y$, as can
also be seen since $Tr(Q)$ and $Tr(Y)$ are nonzero for the $\chi=R$ multiplets
in eq. (\ref{1detc}). We now specialize again to $N_{TC}=2$ so $N_{ETC}=14$.

The requirement that ETC gauge bosons transform SM fermions to the
SM-nonsinglet technifermions and back in order to produce SM fermion masses
entails the following transitions:
\beqs
& & u^{aj}_\chi \to F^{1/2 \ \tau}_\chi + V^{aj}_\tau \cr\cr
& & d^{aj}_\chi \to F^{-1/2\ \tau}_\chi + V^{aj}_\tau \cr\cr
& & \nu^j_\chi  \to F^{1/2 \ \tau}_\chi + U^j_\tau    \cr\cr
& & e^j_\chi    \to F^{-1/2\ \tau}_\chi + U^j_\tau \ . 
\label{trans}
\eeqs
Under ${\rm SU}(2)_{TC} \times {\rm SU}(3)_{gen.} \times {\rm SU}(3)_c \times
{\rm U}(1)_{B-L}$, the $V^{aj}_\tau$ transform as $(2,3,3)_{1/3}$ and the
$U^j_\tau$ as $(2,3,1)_{-1}$, with corresponding electric charges $Q_V=1/6$ and
$Q_U=-1/2$.  To yield the correct generational scales for the SM fermion
masses, the ETC vector boson mass eigenstates should have masses of order
$\Lambda_3 \simeq$ few TeV for $j=3$, $\Lambda_2 \simeq 10^2$ TeV for $j=2$,
and $\Lambda_1 \simeq 10^3$ TeV for $j=1$.  There are also TC-singlet ETC gauge
bosons (i) $X^{aj}_k$ transforming as $(1,8,3)_{4/3}$ with $Q_X=2/3$ involved
in the transitions $u^{aj}_\chi \to \nu^k_\chi + X^{aj}_k$ and $d^{aj}_\chi \to
e^k + X^{aj}_k$; and (ii) $G^j_k$ transforming as $(1,8,1)_0$ occur in the
transitions $f^j_\chi \to f^k_\chi + G^j_k$, where $f=u,d,e,\nu$ and $j,k$ are
generation indices. The ETC gauge bosons contain a subset corresponding to
generators of the Cartan subalgebra of SU(14)$_{ETC}$, which are particle- and
flavor-diagonal; these are generically denoted $V_{dp}$, where $d$ denotes
``diagonal'' and $p=1..., 13$.  

With the fermion content in eq. (\ref{1detc}), the ETC model is vectorial (and
asymptotically free), so that by itself, as the energy scale decreased from
large values, the ETC coupling $\alpha_{ETC}$ would eventually get sufficiently
large to form bilinear fermion condensates, but these would be invariant under
the SU(14)$_{ETC}$ symmetry, which would thus not self-break. To obtain the
sequential dynamical breaking of SU(14)$_{ETC}$ and resultant generational
hierarchy of SM fermion masses, one can augment the model with three auxiliary
strongly coupled gauge symmetries and an appropriately chosen set of chiral
fermions, as in Ref. \cite{ae}.  For our models we would further augment this
with either of the metacolor sectors (1.) or (2.) discussed above.

The flavor-diagonal ETC gauge bosons $V_{dp}$ produce additional contributions
to $(\Delta S)^{(TC)}$ and $(\Delta T)^{(TC)}$ via nondiagonal propagator
corrections in which $Z$ goes to a loop of virtual fermions $\bar f f$ which
then go to $V_{dp}$.  Diagonalizing the vector boson mixing matrix, one finds
that the mass of the physical $Z$ is reduced \cite{zzprime}.  Since
\beq
m_Z^2 = (m_Z^2)_{SM} \, \frac{1-\rho}{1-(m_Z^2)_{SM}G_FS/(2^{3/2}\pi)} \ , 
\label{zmsq}
\eeq
this reduction involves negative and positive contributions to $S$ and $T$,
respectively, which depend on the breaking of SU(14)$_{ETC}$ and resultant
values of $V_{dp}$ masses.

The electrically charged ETC gauge bosons couple directly to the $Z$ via the
$J_{em}$ part of $J_Z = J_{3L}- \sin^2\theta_W J_{em}$ and hence lead to loop
corrections to the $ZZ$ and $Z\gamma$ (and $\gamma\gamma$) 2-point functions.
In contrast to fermion loop corrections, these are gauge-dependent and require
one also to consider non-oblique box and vertex graphs to the same order (as is
the case with analogous $W$ corrections to vector 2-point functions in the SM
\cite{dks}), so that their effects cannot be subsumed into shifts of the
oblique parameters $S$, $T$, and $U$.  The most important corrections involve
the charged ETC vector bosons with lowest masses, $\sim \Lambda_3$.  Because
the ETC gauge bosons are SU(2)$_L$ singlets (cf. eq. (\ref{etcsu2lr})), they do
not couple directly to $W$.

The most important ETC corrections to $BR(Z \to b \bar b)$ arise from graphs in
which the $Z$ produces (i) a virtual $b \bar b$ pair which exchange a $V_{dp}$
with mass $\sim \Lambda_3$ or (ii) a virtual $F^\pm F^\mp$ pair which exchange
a $V^{a3}_\tau$ (also with mass $\Lambda_3$), yielding the outgoing $b \bar b$.
These are analogous to the $V_{d3}$ and $V^3_\tau$ exchanges in a one-family
ETC model, which were found to tend to cancel each other and hence give
acceptably small corrections to this branching ratio \cite{ztobbar}.

With regard to the SM-singlet, TC-nonsinglet fermion sector, it is interesting
to recall that in modern detailed studies of one-family ETC models
\cite{at94}-\cite{kt}, the SM-singlet, ETC-nonsinglet fermion sectors play a
crucial role in the sequential ETC symmetry breaking, and in certain cases
(e.g., for the breaking sequence $G_b$ in \cite{nt} and $S2$ in \cite{ckm,kt}),
they yield SM-singlet sectors of the resultant technicolor field theories that
contain more than just a single right-handed technineutrino $N_R$.  These
studies thus provide explicit examples of how non-minimal SM-singlet
technifermion sectors can arise from ETC breaking. 

\section{Models Having SM-Nonsinglet Technifermions in 
Rank-2 Tensor Representations of SU($N_{TC}$)}

We next discuss the technicolor model of eq. (\ref{1d}) with ${\cal R}_{TC}$
being the symmetric ($S_2$) or antisymmetric ($A_2$) rank-2 tensor
representation of SU($N_{TC}$).  Technifermions in higher-dimensional
representations of $G_{TC}$ have been of interest \cite{higherrep,sann12,sann3}
for several reasons, including walking and the minimization of $S$ (as well as
formal connections with supersymmetric nonabelian gauge theories
\cite{asv},\cite{sann12,sann3}).  
Here they will provide a comparison with our technicolor
models presented in section III with respect to predicted $S$ values and
embedding in ETC.  We first review some of their properties.

We denote the SM-nonsinglet technifermions as $F^{\pm 1/2 \ \tau \tau'}_\chi$,
where $\chi=L,R$, with $F=S_2,A_2$.  The dimensionalities of the rank-$n$
symmetric (antisymmetric) tensor representations of SU($N$) are
$(1/n!)\Pi_{j=0}^{n-1}(N \pm j)$, respectively, so in the TC models of interest
here, there are $d_{S_2,A_2}=(1/2)N_{TC}(N_{TC} \pm 1)$ SU(2)$_L$ doublets
comprised of technifermions.  In cases where $d_{S_2,A_2}$ is odd, one must add
an odd number of other SU(2)$_L$ doublets to avoid a Witten $\pi_4$ anomaly in
the SU(2)$_L$ theory.  Minimally, one would add a single such doublet, and thus
the set of new leptons \cite{sann12} ${\ell^{1/2} \choose \ell^{-1/2}}_L$ and
$\ell^{\pm 1/2}_R$.  The $\ell^{\pm 1/2}$ must get masses that are sufficiently
large, $\gsim 100$ GeV, to have escaped detection.  This addition is necessary,
for example, in the case $N_{TC}=2$, $F=S_2$, where $d_{S_2}=3$.  The models
with $F=S_2$ and $N_{TC}=2$, and possibly also $N_{TC}=3$, could plausibly
exhibit walking \cite{sann12}.  For the $N_{TC}=2$ case, owing to the necessity
of adding the new heavy lepton SU(2)$_L$ doublet, the total new physics
contribution to $S$ is comprised of the three technifermion SU(2)$_L$ doublets
and the heavy lepton doublet, so that $(\Delta S)^{(NP)}_{pert.} = 2/(3\pi)$.
This is larger by a factor of 2 than the value in our models, given in
eq. (\ref{deltas_1d}). The full nonperturbative $(\Delta S)^{(NP)}$ values
involve walking reductions (relative to the respective $\sim 0.1N_D$
non-walking estimates).  In these models with higher technifermion
representations where the walking occurs with the given SM-nonsinglet
technifermions, one would not add SM-singlet technifermions.  Values of
$N_{TC}$ higher than 2 yield larger values of $(\Delta S)^{(TC)}$ and hence are
less well motivated.

Regarding the $F=A_2$ case, we first observe that for $N_{TC}=3$, this
antisymmetric rank-2 tensor degenerates to just the $\bar{\underline{3}}$
(conjugate fundamental) representation, for which $N_{F,c} \simeq 12$ (cf. 
eq. (\ref{nfca}).  Since
eq. (\ref{1d}) corresponds to the substantially smaller value, $N_F=2$, this
case would not be expected to exhibit walking, so that $(\Delta
S)^{(TC)}_{pert.} = 1/(2\pi)$ without a walking reduction.  Higher values of 
$N_{TC}=4$ yield larger values of $(\Delta S)^{(TC)}_{pert.}$.

\section{Embedding of TC Models with ${\cal R}_{TC}=S_2,A_2$ in ETC} 

In this section we investigate embeddings of technicolor models with ${\cal
R}_{TC}=S_2,A_2$ in an ETC theory.  For generality, we will usually take
$N_{TC}$ to be arbitrary. 

\subsection{One-Doublet TC}

We consider first the case where the technifermions are described by
eq. (\ref{1d}), so that they have the explicit form
\beq
{F^{1/2 \ \tau \tau'} \choose F^{-1/2 \ \tau \tau'}}_L \ , \quad 
F^{\pm 1/2 \ \tau\tau'}_R \ . 
\label{1dhr}
\eeq
In constructing the high-energy ETC-symmetric theory, one treats all of the ETC
indices on an equal footing, so the natural embedding of the technifermions in
eq. (\ref{1dhr}) would be a rank-2 symmetric or antisymmetric representation of
${\rm SU}(N_{TC}+12)$. But this is excluded since, among other things, it would
lead to various light leptoquark fermions with SM quantum numbers given by
$\underline 3$ of SU(3)$_c$, with lepton number $L=1$ (and generational indices
$jk$), which are not observed experimentally.  (For $F=S_2$, it would also
imply fermions transforming as \underline{6}'s of SU(3)$_c$ (``quixes''); for
$F=A_2$, it would imply SU(2)$_L$ doublets of $\underline{\bar 3}$'s of
SU(3)$_c$, etc.)

In view of this negative result, one is motivated to investigate whether a
higher-dimensional representation of the technicolor group SU($N_{TC}$) could
occur in a fundamental representation of the extended technicolor group
SU($N_{ETC}$) which contains SU($N_{TC}$).  This does not occur for regular
embeddings of ${\rm SU}(N_{TC}) \subset {\rm SU}(N_{ETC})$.  (Here, a regular
embedding of a subgroup $H$ in a Lie group $G$ is one in which the generators 
of the Lie algebra of $H$ can be written as a restriction of, or subset of, the
generators of the Lie algebra of $G$.)  In contrast, for embeddings of
subgroups $H \subset G$ which are not of this type (and are called ``special''
embeddings \cite{slansky}), it is possible for the fundamental representation
of a Lie group $G$ to decompose, with respect to a subgroup $H$ in such a
manner as to yield a higher-dimensional representation of $H$.  For example,
with a special embedding of SU(2) in SU(3), the decomposition of the
$\underline 3$ of SU(3) yields a $\underline 3$ of SU(2) \cite{slansky}.
However, we have not found any cases that appear promising for semirealistic
(E)TC models with $F=S_2,A_2$.  It thus remains a challenge to construct
acceptable ETC models that yield TC sectors with higher-dimensional
technifermion representations.

\subsection{One-Family TC} 

Among models with higher TC representations, the minimization of $(\Delta
S)^{(TC)}$ motivates one to focus on the one-doublet case.  Nevertheless, it is
of some interest to consider how one would try to embed a one-family TC model
with $F=S_2$ or $F=A_2$ in ETC.  This provides a different perspective
on how generations might arise, although, as we shall show, it fails to yield
an acceptable TC theory.  

First, recall, as background, how this embedding is carried out for the simpler
case of a one-family SU($N_{TC}$) model with ${\cal R}_{TC}=fund.$.  In both
cases, $[G_{ETC},G_{SM}]=0$ and ${\rm SU}(N_{TC}) \subset {\rm SU}(N_{ETC})$.
For general $N_{TC}$, the technifermions are ${U^{a \tau} \choose D^{a
\tau}}_L$, $U^{a \tau}_R$, $D^{a \tau}_R$, ${N^\tau \choose E^\tau}_L$,
$N^\tau_R$, and $E^\tau_R$. One forms the ETC multiplets with these SM
transformation properties by gauging the generation index and combining it with
the technicolor index, so that $N_{ETC}=N_{gen.}+N_{TC}=3+N_{TC}$.  Thus, with
the minimal value $N_{TC}=2$, the ETC group would be SU(5)$_{ETC}$ and, for
example, the ETC multiplet transforming as $(5,3,1)_{4/3,R}$ under ${\rm
SU}(5)_{ETC} \times {\rm SU}(3)_c \times {\rm SU}(2)_L \times {\rm U}(1)_Y$
would be $(u,c,t,U^4,U^5)_R$.  (For $N_{TC}=2$, one can also construct ETC
models with some fermions being assigned to conjugate fundamental
representations \cite{ckm,kt}.)

For the cases $F=S_2,A_2$ we first determine $N_{ETC}$ for a given
$N_{TC}$.  Let
\beq
N_{ETC} = m + N_{TC}
\label{nrel2}
\eeq
so that 
\beq
{\rm SU}(N_{ETC}) \supset {\rm SU}(m) \times {\rm SU}(N_{TC}) \ . 
\label{ellxtc}
\eeq
We next extend the $S_2$ and $A_2$ representations of SU($N_{TC}$) to
corresponding representations of SU($N_{ETC}$), denoted with the same
symbols. With respect to the direct product subgroup (\ref{ellxtc}) these
transform as follows:
\beq
S_2: \ ( \frac{m(m+1)}{2},1 ) + (m,N_{TC}) + (1,\frac{N_{TC}(N_{TC}+1)}{2})
\label{sreps}
\eeq
\beq
A_2: \ ( \frac{m(m-1)}{2},1 ) + (m,N_{TC}) + (1,\frac{N_{TC}(N_{TC}-1)}{2})
\label{areps}
\eeq
Thus, for an $S_2$ or $A_2$ ETC fermion multiplet transforming
according to a given representation of $G_{SM}$, the number of technisinglet
components, which should be equal to the number of generations, is
\beq
(N_{gen.})_{S_2,A_2} = \frac{m(m \pm 1)}{2} + \delta_{A_2;N_{TC}=2} \ , 
\label{ngensa}
\eeq
where the second term is a Kronecker delta function which is equal to one if
$F=A_2$ and $N_{TC}=2$ and zero otherwise.  This second term is present because
in eq. (\ref{areps}), if $N_{TC}=2$, the third representation is (1,1), a
technisinglet.  The first few sets of pairs for $F=S_2$ are $(m,N_{gen.})$ are
(1,1), (2,3), and (3,6), so that in order to reproduce the physical value,
$N_{gen.}=3$, one would take $m=2$, whence
\beq
N_{ETC} = 2 + N_{TC} \quad {\rm for} \ \ F=S_2 \ . 
\label{netc_s}
\eeq
The technisinglet components are then
$(\psi^{11}_\chi,\psi^{12}_\chi,\psi^{22}_\chi)$, $\chi=L,R$.  Parenthetically,
we note that with the $F=S_2$ assignment, among toy-model values of $N_{gen.}$,
one could accomodate $N_{gen.}=1$ (for $m=1$), but not $N_{gen.}=2$.

For the case $F=A_2$, the first few $(m,N_{gen.})$ values are (i) (1,1), (2,2),
(3,4), etc. for $N_{TC}=2$; (ii) (1,0), (2,1), (3,3), etc. for $N_{TC} \ge 3$.
Evidently, the model with $F=A_2$ and $N_{TC}=2$ is not able to accomodate
three SM fermion generations, while for $N_{TC} \ge 3$, this is possible with
$m=3$, so
\beq
N_{ETC} = 3+N_{TC} \quad {\rm for} \ \ F=A_2, \ N_{TC} \ge 3 \ . 
\label{netc_a}
\eeq
In this case, the three generations of a (technisinglet) fermion field with a
given set of SM quantum numbers can be written as
$(\psi^{23}_\chi,\psi^{31}_\chi,\psi^{12}_\chi)$, $\chi=L,R$, where the order
is a convention.  For this $F=A_2$ case one would preferentially choose the
minimal possible value, $N_{TC}=3$ to minimize technicolor contributions to the
electroweak $S$ parameter.

The fact that there are restrictions on the possible values of $N_{gen.}$ in
these TC models with either $S_2$ or $A_2$ fermions is quite different from the
situation in ETC models in which the SM-nonsinglet fermions transform according
to the fundamental representation of SU($N_{ETC}$) and where one can accomodate
an arbitrary number of SM fermion generations, subject to the constraints of
asymptotic freedom of the color and technicolor groups.

We next calculate the leading-order term in the technicolor beta function.
Using the values of $d_{S_2}$ and $d_{A_2}$ and the fact that there are
$N_w(N_c+1)=8$ Dirac fermion components for each TC gauge index, we have, for
the one-family model,
\beqs
b^{(TC)}_0 & = & -\frac{1}{3}[5N_{TC} + 16(m \pm 2) \cr\cr
                & + & 2\sum_{SMS \ f} T({\cal R}_{TC,f}) ] \ , 
\label{b0gentc}
\eeqs
where the $+$ and $-$ signs apply for $F=S_2,A_2$, respectively, and the last
term is the contribution from possible SM-singlet (SMS) fermions.  For both of
the relevant cases (i) $F=S_2$ and hence $m=2$; and (ii) $F=A_2$ and hence
$m=3$, $N_{TC} \ge 3$, $b^{(TC)} < 0$, i.e., the technicolor theory is not
asymptotically free.  Since asymptotic freedom is a necessary property of the
technicolor theory, being responsible for the confinement and formation of the
technifermion condensates that break the electroweak gauge symmetry, this lack
of asymptotic freedom rules out these models. 

\section{Conclusions}

In this paper we have studied fermion representations in technicolor theories
with the goal of minimizing technicolor corrections to precision electroweak
quantities, in particular, the $S$ parameter.  We have constructed SU(2)$_{TC}$
models with standard-model nonsinglet technifermion sectors of the form of
eq. (\ref{1d}) with $I_{TC}=1/2$ technifermions which also plausibly have the
desirable property of walking behavior, owing to SM-singlet technifermion
sectors.  As a consequence, these models yield the rather small estimate
$(\Delta S)^{(TC)}_{pert.}$ in eq. (\ref{deltas_1d}), and a full
(nonperturbatively calculated) $(\Delta S)^{(TC)}$ which is reduced by the
walking.  We have contrasted our results with some models that obtain walking
via technifermions in higher-dimensional representations of the technicolor
group.  For both of these types of models, we have analyzed embeddings in
extended technicolor theories.  The attractively small value of $(\Delta
S)^{(TC)}$ in the one-doublet walking technicolor models with ${\cal
R}_{TC}=fund.$ that we have discussed motivates further study of their
embeddings in ETC and the resultant phenomenological predictions, in 
particular, the differences with respect to one-family ETC models.

\bigskip

\acknowledgements

This research was partially supported by the grant NSF-PHY-00-98527.  We thank
T. Appelquist, M. Kurachi, and F. Sannino for useful discussions. 

\section{Appendix} 

We include here some relevant results used in the text. The beta function for
a given gauge interaction $G_j$ is
\beq
\beta_j = \frac{d \alpha_j}{dt} = - \frac{\alpha_j^2}{2\pi}\left ( b_0^{(j)} +
\frac{b_1^{(j)}}{4\pi}\alpha_j + O(\alpha_j^2) \right ) \ ,
\label{beta}
\eeq
where $\alpha_j =g_j^2/(4\pi)$, $t=\ln \mu$, and the first two terms
$b_0^{(j)}$ and $b_1^{(j)}$ are scheme-independent.  Provided that $b_0^{(j)} >
0$, i.e., the theory is asymptotically free, there is an infrared-stable fixed
point of the renormalization group equation if $b^{(j)}_1 < 0$, at
$\alpha_{j,IR} = -4 \pi b^{(j)}_0/b^{(j)}_1$.  

Now let $G_j=G_{TC}$. For a technifermion transforming according to a
representation ${\cal R}_{TC}$ of $G_{TC}$, the critical value of $\alpha_{TC}$
for which a bilinear technifermion condensate forms is denoted $\alpha_{TC,c}$.
An analysis of the Schwinger-Dyson gap equation yields the estimate
\cite{wtc,gapeq} $\alpha_{TC,c} \simeq \pi/(3 C_2({\cal R}_{TC}))$, where
$C_2({\cal R})$ is defined by $\sum_{a=1}^{order(G)} \sum_{j=1}^{dim({\cal R})}
(T_a)_{ij}(T_a)_{jk} = C_2({\cal R})\delta_{ik}$.  (We also define $T({\cal
R})$ via $\sum_{i,j=1}^{{\rm dim}({\cal R})}(T_a)^i_j \ (T_b)^j_i = T({\cal R})
\delta_{ab}$.)  This estimate of $\alpha_{TC,c}$ involves some theoretical
uncertainty because of the strong coupling involved.  A vectorial SU($N_{TC}$)
theory with $N_f$ technifermions in the fundamental representation is expected
to exist in a confining phase with S$\chi$SB if $N_f < N_{f,c}$, where
\cite{chiralpt}
\beq
N_{f,c} \simeq \frac{2N_{TC}(50N_{TC}^2-33)}{5(5N_{TC}^2-3)}
\label{nfca}
\eeq
and in a nonabelian Coulomb phase if $N_{f,c} < N_f < 11N_{TC}/2$.  For $N_{TC}
= 2$ and $N_{TC}=3$ we have $N_{f,c} \simeq 8$ and  $N_{f,c} \simeq 12$,
respectively. 

In the part of the text dealing with higher-representation technifermions, the
motivation was their effect on walking.  Here we comment parenthetically on a
different application of higher-representation technifermions, namely the idea
of using these higher representations to produce the generational mass
hierarchy for the SM fermions.  Thus, consider, for example, a (vectorial)
technicolor theory with a set of SM-nonsinglet technifermions given by
eq. (\ref{1d}) with three different ${\cal R}_{TC,j}$'s such that $dim({\cal
R}_{TC,j})$ is an increasing function of the generation index $j$.  Provided
that the technicolor theory is asymptotically free and in the phase with
spontaneous chiral symmetry breaking (instead of a possible nonabelian Coulomb
phase), as the energy scale decreases through the TeV region, there is a
hierarchy of scales at which the technifermions of different representations
condense, and hence a hierarchy of dynamical technifermion masses
$\Sigma_{TC,j}$.  If one could arrange the ETC dynamics so that, to leading
order, the SM fermions of generation $j$ communicate with the technifermions
with TC representation ${\cal R}_{TC,j}$, then the resultant masses of the SM
fermions of generation $j$, namely $m_{f_j} \propto \eta_j
\Sigma_{TC,j}^3/\Lambda_{ETC,j}^2$ (where $\eta_j$ is a possible walking
factor), could exhibit a hierarchy due to a combination of the hierarchies in
$\Sigma_{TC,j}$ and $\Lambda_{ETC,j}$, in contrast to the situation in usual
ETC models with only one $\Sigma_{TC}$ scale.  For the technifermions in
eq. (\ref{1d}) with ${\cal R}_{TC,j}$, one has $(\Delta S)^{(TC)}_{pert.} =
(6\pi)^{-1}\sum_{j=1}^3 dim({\cal R}_{TC,j})$.  Examination of specific models
shows that the resultant values of $(\Delta S)^{(TC)}$ are excessively
large. It also appears difficult to construct models with the appropriate ETC
dynamics.

\end{document}